\documentclass[showpacs,amsmath,amssymb,epsfig]{revtex4}

\usepackage{graphicx}
\usepackage{dcolumn}
\usepackage{bm}

\begin{document}

\title{Higher Twist Analysis of the Proton $g_1$ Structure Function}

\author{M.~Osipenko$^1$, W.~Melnitchouk$^2$, S.~Simula$^3$,
	P.~Bosted$^2$, V.~Burkert$^2$, M.~E.~Christy$^4$,
	K.~Griffioen$^5$, C.~Keppel$^{2,4}$, S.~E.~Kuhn$^6$}
\affiliation{
	$^1$ INFN, Sezione di Genova, 16146, Genoa, Italy	\\
	$^2$ \mbox{Jefferson Lab, 12000 Jefferson Avenue,
		Newport News, Virginia 23606, USA}		\\
	$^3$ INFN, Sezione Roma III, 00146 Roma, Italy		\\
	$^4$ Hampton University, Hampton, Virginia 23668, USA	\\
	$^5$ \mbox{College of William \& Mary,
		Williamsburg, Virginia, 23187, USA} \\
	$^6$ \mbox{Old Dominion University,
		Norfolk, Virginia 23529, USA}}

\begin{abstract}
We perform a global analysis of all available spin-dependent proton
structure function data, covering a large range of $Q^2$,
$1 \leq Q^2 \leq 30$~GeV$^2$, and calculate the lowest moment of the
$g_1$ structure function as a function of $Q^2$.
From the $Q^2$ dependence of the lowest moment we extract matrix
elements of twist-4 operators, and determine the color electric
and magnetic polarizabilities of the proton to be
$\chi_E =  0.026\ \pm\ 0.015\ ({\rm stat.})\
		  \pm\ {}^{0.021}_{0.024}\ ({\rm sys.})$
and
$\chi_B = -0.013\ \mp\ 0.007\ ({\rm stat.})\ 
		  \mp\ {}^{0.010}_{0.012}\ ({\rm sys.})$,
respectively.
\end{abstract}


\maketitle

%%%%%%%%%%%%%%%%%%%%%%%%%%%%%%%%%%%%%%%%%%%%%%%%%%%%%%%%%%%%%%%%%%%%%%%%%
Measurements of spin-dependent structure functions of the proton
reveal fundamental information about the proton's quark and gluon
structure.
In the quark-parton model, the $g_1$ structure function is interpreted
in terms of distributions of quarks carrying light-cone momentum
fraction $x$, with spins aligned versus anti-aligned with that of the
nucleon.
The lowest moment, or integral over $x$, of $g_1$ also determines the
total spin carried by quarks in the nucleon.

Although most structure function studies have focused on the scaling
regime at high four-momentum transfer squared, $Q^2$, the behavior
of $g_1$ and its moments in the transition region at intermediate
$Q^2$ ($\sim 1$~GeV$^2$) can reveal rich information about the
long-distance structure of the nucleon.
One example of the complexity of this region is the transition from
the Bjorken or Ellis-Jaffe sum rules at high $Q^2$ to the
Gerasimov-Drell-Hearn sum rule at $Q^2=0$ \cite{BI}.

Of particular importance in this region is the role of the nucleon
resonances, and the interplay between resonant and scaling
contributions.
According to the operator product expansion (OPE) in QCD, the
appearance of scaling violations at low $Q^2$ is related to the
size of higher twist corrections to moments of structure functions
\cite{DGP}.
Higher twists are expressed as matrix elements of operators involving
nonperturbative interactions between quarks and gluons.
The study of higher twist corrections thus gives us direct insight
into the nature of long-range quark-gluon correlations.

In this paper we determine the size of the higher twist contributions
to the lowest moment of the $g_1$ structure function of the proton.
We analyze the entire set of available data from experiments at
SLAC \cite{SLAC-E80,SLAC-E130,SLAC-E143,SLAC-E155},
CERN \cite{EMC-NA2,SMC-NA47}, DESY \cite{HERMES},
and most recently Jefferson Lab \cite{CLAS},
where high-precision data in the resonance region and at low and
intermediate $Q^2$ have been taken.
Combining the moment data from the various experiments is nontrivial,
however, since different analyses typically make use of different
assumptions about extrapolations into unmeasured regions of kinematics.
In the present analysis, we therefore extract the structure function
moment using a single set of inputs and assumptions for {\em all}
the data.

The lowest (Cornwall-Norton) moment of the proton $g_1$ structure
function is defined as
\begin{equation}
\label{eq:gam1}
\Gamma_1(Q^2)
= \int_0^1 dx\ g_1(x,Q^2)\ .
\end{equation}
The upper limit includes the proton elastic contribution at
$x \equiv Q^2/2M\nu = 1$, where $\nu$ is the energy transfer,
and $M$ is the proton mass.
The inclusion of the elastic component is essential if one wishes to
use the OPE to study the evolution of the integral in the moderate
$Q^2$ region \cite{JU_G1}.

From the OPE, at large $Q^2$ the moment $\Gamma_1$ can be expanded
in powers of $1/Q^2$, with the expansion coefficients related to
nucleon matrix elements of operators of a definite twist
(defined as the dimension minus the spin of the operator).
At high $Q^2$ the moment is dominated by the leading twist contribution,
$\mu_2$, which is given in terms of matrix elements of the twist-2 axial
vector current, $\bar\psi\ \gamma^\mu \gamma_5\ \psi$.
This can be decomposed into flavor singlet and nonsinglet
contributions as
\begin{eqnarray}
\label{eq:mu2}
\mu_2(Q^2)
&=& C_{\rm s}(Q^2) {a_0^{\rm inv} \over 9}\
 +\ C_{\rm ns}(Q^2) \left( {a_3 \over 12}\ +\ {a_8 \over 36} \right)
\end{eqnarray}
where $C_{\rm s}$ and $C_{\rm ns}$ are the singlet and nonsinglet
Wilson coefficients, respectively \cite{LARIN}, which are calculated
as a series in $\alpha_s$.
The triplet and octet axial charges, $a_3 = g_A = 1.267$ and
$a_8 = 0.58$, are extracted from weak decay matrix elements.
For the singlet axial charge, we work with the renormalization
group invariant definition in the $\overline{\rm MS}$ scheme,
$a_0^{\rm inv}=a_0(Q^2=\infty)$, in which all of the $Q^2$
dependence is factorized into the Wilson coefficient $C_{\rm s}$.

Considerable effort has been made over the past two decades in
determining the singlet axial charge, which in the quark-parton
model is identified with the total spin carried by quarks in the
proton.
In this work we focus instead on using the world's data to extract
the coefficient of the $1/Q^2$ subleading, twist-4 term, which
contains information on quark-gluon correlations in the nucleon.

In addition to the twist-4 matrix element, the $1/Q^2$ term also
contains so-called ``kinematic'' higher twists, associated with
target mass corrections (which are formally twist-2), and the $g_2$
structure function, which is obtained from measurements with
transversely polarized targets.
One technique for removing these from the $1/Q^2$ correction
is to work in terms of the Nachtmann moment \cite{NACHT_POL},
\begin{eqnarray}
\label{eq:M1}
M_1(Q^2)
&=& \int_0^1 dx\ {\xi^2 \over x^2}
    \left\{ g_1(x,Q^2) \left( {x \over \xi}
			    - {1 \over 9}{M^2 x \xi \over Q^2}
		       \right)
	  - g_2(x,Q^2) {4 \over 3} {M^2 x^2 \over Q^2}
    \right\}\ ,
\end{eqnarray}                                                            
where $\xi = 2x / (1 + \sqrt{1 + 4 M^2 x^2/Q^2})$ is the Nachtmann
scaling variable.
The twist expansion of $M_1(Q^2)$ then yields
\begin{eqnarray}
\label{eq:M1twist}
M_1(Q^2)
&=& \mu_2(Q^2) + { \mu_4(Q^2) \over Q^2 }
 + { \mu_6(Q^2) \over Q^4 } + \cdots
\end{eqnarray}
where $\mu_2$ is given in Eq.~(\ref{eq:mu2}).
The $1/Q^2$ correction in Eq.~(\ref{eq:M1twist}) exposes directly
the ``dynamical'' twist-4 coefficient $f_2$, since
$\mu_4(Q^2) = 4 f_2(Q^2) / 9 M^2$, where $f_2$ is given in terms
of a mixed quark-gluon operator,
\begin{eqnarray}
\label{eq:f2op}
f_2(Q^2)\ M^2 S^\mu
&=& {1 \over 2} \sum_q e_q^2\
    \langle N |
	g\ \bar\psi_q\ \widetilde{G}^{\mu\nu} \gamma_\nu\
	\psi_q
    | N \rangle .
\end{eqnarray}
Here $\widetilde G^{\mu\nu}
     = {1\over 2} \epsilon^{\mu\nu\alpha\beta} G_{\alpha\beta}$
is the dual of the gluon field strength tensor,
$S^\mu$ is the proton spin vector, $g$ is the strong coupling
constant, and $e_q$ is the quark charge.

Clearly the $1/Q^2$ term can be best determined in the intermediate
$Q^2$ region, where $Q^2$ is neither so large as to completely
suppress the higher twists, nor so small as to render the twist
expansion unreliable.
A meaningful analysis of data from different experiments further
requires that the same set of inputs be used in the determination of
$g_1$, as well as $g_2$.

In practice one must reconstruct $g_1$ from a combination of
longitudinal ($A_\parallel$) and transverse ($A_\perp$) polarization
asymmetries, together with the unpolarized $F_1$ structure function,
and the ratio $R$ of the longitudinal to transverse cross sections.
We begin by collecting all available data on $A_\parallel$,
as published in 
Refs.~\cite{SLAC-E80,SLAC-E130,SLAC-E143,SLAC-E155,EMC-NA2,SMC-NA47,HERMES,CLAS},
and use the same inputs in the analysis of $A_\perp$, $F_1$ and $R$
for all the data sets.

To provide a realistic description of $A_\perp$, or equivalently
$A_2$ (which is given in terms of $A_\parallel$ and $A_\perp$
\cite{SLAC-E143}), we consider both the resonance and nonresonant
background contributions.
For the background we use the (twist-2) Wandzura-Wilczek (WW) relation
\cite{WW}.
Inclusion of target mass corrections in the WW formula \cite{WW}
enables this prescription to be used down to low $Q^2$, where
target mass corrections are known to be important \cite{DGP}.
In the resonance region, however, the WW approximation fails,
and here alternative parameterizations are required.
We calculate the resonance contribution from the electromagnetic
helicity amplitudes $S_{1/2}(Q^2)$ and $A_{1/2}(Q^2)$ obtained in the
constituent quark model \cite{GIANNINI}, which includes 14 major
resonances.
The resonance contribution is then normalized to satisfy
the Burkhardt-Cottingham sum rule \cite{BC}.
The $A_2$ model is consistent with the available data
\cite{SLAC-E155X,A2_OTHER}, as well as with the model-independent
Soffer bound \cite{SOFFER}.

\begin{figure}[t]
\begin{center}
\hspace*{1cm}
\includegraphics[bb=2cm 2cm 22cm 24cm, scale=0.4]{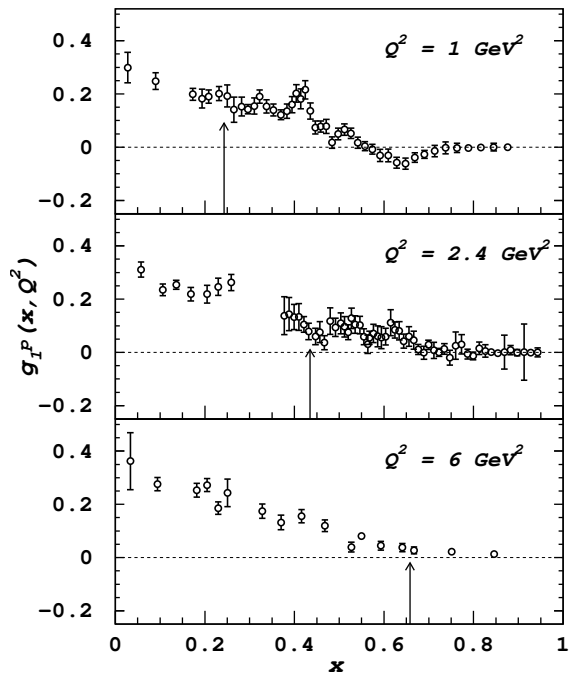}
\caption{Proton $g_1$ structure function at several different
        $Q^2$ values.  The points represent reanalyzed experimental
        data obtained from the longitudinal asymmetry $A_\parallel$
        from Refs.~\cite{SLAC-E80,SLAC-E130,SLAC-E143,SLAC-E155,EMC-NA2,SMC-NA47,HERMES,CLAS}
        using the procedure described in the text.
        The vertical arrows indicate the boundary between the
        resonance (to the right of the arrow) and deep inelastic regions
        ($W=2$~GeV).}
\end{center}
\label{fig:1}
\end{figure}

For the ratio $R(x,Q^2)$ we use a new parametrization based on
Rosenbluth-separated cross sections \cite{R_PARAM,BIGPAPER},
which is adapted to the low-$Q^2$ and low-$W^2$ region, and smoothly
interpolates to the earlier parameterization of the deep inelastic
region from Ref.~\cite{R1998}.
This parameterization uses all published data in the resonance region
\cite{RES_R}, as well as new data from Ref.~\cite{R_PARAM}.

The spin-averaged proton cross section is well determined in both
the resonance and deep inelastic regions.
At high $Q^2$ the effect of $R$ is small and the differential
cross section $d\sigma/d\Omega dE^\prime$ is proportional to $F_2$
(or $F_1$).
At moderate $Q^2$, however, the extraction of $F_1$ and $F_2$ from
the cross section requires knowledge of $R$.
Using the parametrization of $R$ from Ref.~\cite{R_PARAM}, we
constructed a database of world data on $F_1$ from which values of
$F_1$ corresponding to the measured $A_\parallel$ kinematic points
were obtained by interpolation.
Most of the data points for $A_\parallel$ have kinematically close
sets of $d^2\sigma/d\Omega dE^\prime$ points, which allows
interpolation uncertainties to be minimized.
Full details of the extraction of $g_1$ will be provided in a
forthcoming publication \cite{BIGPAPER}.

The resulting $g_1$ structure function is illustrated in Fig.~1
as a function of $x$ for several representative $Q^2$ values.
The vertical arrows indicate the boundary between the resonance and  
deep inelastic regions at $W=2$~GeV.
At the lower $Q^2$ values, $Q^2 \sim 1$~GeV$^2$, a significant portion
of the $x$ range is in the resonance region (contributing $\sim 40\%$   
of the magnitude of the lowest moment).

\begin{figure}[t]
\begin{center}
\centerline{\includegraphics[scale=0.35,angle=270]{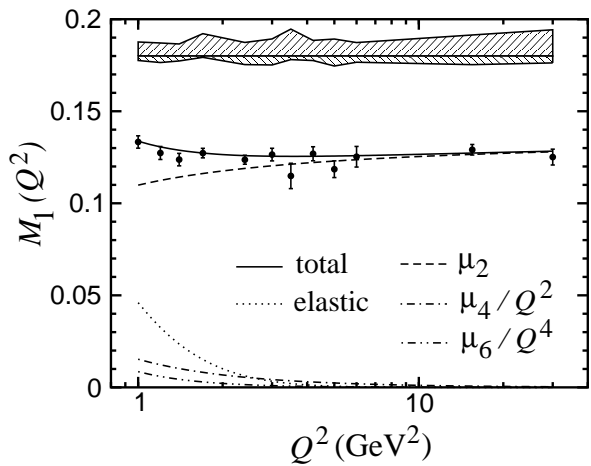}}
\end{center}
\vspace*{-0.5cm}
\caption{$Q^2$ dependence of the Nachtmann moment $M_1(Q^2)$.
	The error bars are statistical, with the systematic errors
	indicated by the hashed areas (see text).
	The leading twist (dashed), $1/Q^2$ (dot-dashed),
	$1/Q^4$ (dot-dot-dashed) and elastic (dotted) contributions
	are shown separately.  The solid curve is the sum of
	leading and higher twist terms.}
\label{fig:2}
\end{figure}

The first moment of $g_1$ is evaluated using the same method as in
recent analyses of the unpolarized proton $F_2$ structure function
moments \cite{OSIPENKO,ARMSTRONG}.
This method is essentially independent of assumptions about the $x$
dependence of the structure function when interpolating between data
points, and is therefore well-suited for a study of $Q^2$ evolution
of the moments.
For the low-$x$ extrapolation, beyond the region where data exist, we
use the Regge model-inspired parametrization from Ref.~\cite{SIMULA}.
To estimate the uncertainty associated with the low-$x$ extrapolation,
we also consider other parameterizations \cite{OTHER}, and take the
maximum difference between the respective low-$x$ contributions as
the error.

The resulting Nachtmann moment $M_1(Q^2)$ is shown in Fig.~2,
where the error bars on the data points are statistical only.
The systematic errors, some of which are correlated, are shown
separately in the hashed areas above the data, and represent
uncertainties from the low-$x$ extrapolation (lower hashed area),
and the experimental systematic errors together with those from
$A_2$, $R$ and an estimated 5\% uncertainty on the elastic
contribution (upper hashed area).
The $g_2$ contribution to $M_1$ is obtained from $A_\parallel$,
$A_\perp$, and $F_1$, as determined from the present analysis
(see Ref.~\cite{BIGPAPER} for details).

The fit to the total moment $M_1(Q^2)$ uses three parameters,
$a_0^{\rm inv}$, $f_2$ (or $\mu_4$) and $\mu_6$, with the non-singlet
axial charges ($g_A$ and $a_8$) as inputs.
For the leading twist contribution we use a next-to-leading order
approximation for the Wilson coefficients and the two-loop expression
for $\alpha_s$, which at $Q^2 = 1$~GeV$^2$ corresponds to
$\alpha_s^{\rm NLO} = 0.45 \pm 0.05$ in the $\overline{\rm MS}$ scheme.

In fitting the parameters, we have considered both multiparameter
(simultaneous) fits and sequential fits, in which the leading twist
term $a_0^{\rm inv}$ is first fitted to the high-$Q^2$ data, and then
the higher twist terms are extracted.
While both methods should in principle yield the same results when
the experimental errors are small, in practice the multiparameter fit
may not be most suitable choice when emphasizing the high-precision
low-$Q^2$ data.
The multiparameter fit is most effective when the errors on the data are
similar across the entire $Q^2$ range, and the number of points in the
region which determines the leading twist contribution
($Q^2 \agt 5$~GeV$^2$) is similar to that which constrains the higher
twists ($Q^2 \alt 5$~GeV$^2$).

Assuming the data at high $Q^2$ are saturated by the twist-2 term
alone, the fit to the $Q^2 > 5$~GeV$^2$ data determines the singlet
axial charge to be
\begin{eqnarray}
a_0^{\rm inv}
&=& 0.145\ \pm\ 0.018\ {\rm (stat.)}\ \pm\ 0.103\ {\rm (sys.)}
\pm\ 0.041\ ({\rm low}\ x)\ \pm\ {}^{0.006}_{0.010}\ (\alpha_s)\ ,
\label{eq:a0}
\end{eqnarray}
where the first and second errors are statistical and systematic,
the third comes from the $x \to 0$ extrapolation, and the last is
due to the uncertainty in $\alpha_s$.
We have considered the sensitivity of the results to the value of
$Q^2$ used to constrain the leading twist term.
We find that $a_0^{\rm inv}$ converges to the above value for
$Q^2 \agt 3$--4~GeV$^2$.
Fitting the $Q^2 > 10$~GeV$^2$ data would lead to practically the
same values of $a_0^{\rm inv}$, but with a slightly larger error bar.

Having determined the twist-2 term from the high-$Q^2$ data,
we now extract the $1/Q^2$ and $1/Q^4$ coefficients from the
$1 \leq Q^2 \leq 5$~GeV$^2$ data, fixing $a_0^{\rm inv}$ to the
above value, but allowing it to vary within its statistical errors.
For the twist-4 coefficient we find
\begin{eqnarray}
f_2
&=& 0.039\ \pm\ 0.022\ {\rm (stat.)}\
	   \pm\ {}^{0.000}_{0.018}\ {\rm (sys.)}
	   \pm\ 0.030\ ({\rm low}\ x)\
    	   \pm\ {}^{0.007}_{0.011}\ (\alpha_s)\ ,
\label{eq:f2}
\end{eqnarray}
normalized at a scale $Q^2=1$~GeV$^2$
(the $Q^2$ evolution of $f_2$ is implemented using the one-loop
anomalous dimensions calculated in Refs.~\cite{SV}).
The systematic uncertainty on $f_2$ is determined by refitting the
$M_1$ data shifted up or down by the $M_1$ systematic uncertainty
shown in Fig.~2 (upper hashed area).
The low-$x$ extrapolation uncertainty is determined by fitting the
$M_1$ values shifted by the maximum difference between the $x \to 0$
contributions calculated with the parameterizations from
Refs.~\cite{SIMULA} and \cite{OTHER} (lower hashed area in Fig.~2).
The relative contribution from the low-$x$ extrapolation to the
asymptotic value of $M_1$ is $\approx 7\%$ and 15\% at $Q^2 = 1$
and 3~GeV$^2$, respectively.
However, because the $x \to 0$ extrapolation affects more the overall
magnitude of $M_1$ rather than its $Q^2$ dependence, the effect on
$f_2$ is relatively small.

For the $1/Q^4$ term, the best fit to the $1 \leq Q^2 \leq 5$~GeV$^2$
data yields
$\mu_6/M^4 = 0.011 \pm 0.013\ ({\rm stat.})
	       \pm {}^{0.010}_{0.000}\ ({\rm sys.})
	       \pm 0.011\ ({\rm low}\ x)
	       \pm 0.000\ (\alpha_s)$,
with the errors determined as for $f_2$.
Within the present level of accuracy the $Q^2$ evolution of the
coefficient $\mu_6$ can be neglected.
The $1/Q^2$ and $1/Q^4$ contributions to $M_1$ are illustrated in Fig.~2.
For comparison, the elastic contribution is also shown, as is the sum of 
the leading plus higher twist contributions.
Since one is fitting $M_1$ in a relatively low $Q^2$ region, one may
ask whether still higher-order corrections could be significant beyond
those considered in Eq.~(\ref{eq:M1twist}).
Adding a phenomenological $\mu_8/Q^6$ term to the $Q^2 > 1$~GeV$^2$
fit, and fixing the other parameters to their quoted values, gives a
coefficient $\mu_8/M^6 = -0.004 \pm 0.004$ which is consistent with
zero.
To determine $\mu_8$ more precisely one needs to go below
$Q^2 \sim 1$~GeV$^2$, however, it is not clear that the twist
expansion is convergent at such low $Q^2$.

Simply fitting the entire $1 < Q^2 < 30$~GeV$^2$ data set using a
3-parameter fit, the value of the singlet charge would be essentually
unchanged ($a_0^{\rm inv} = 0.145 \pm 0.023\ {\rm (stat.)}$), while
the twist-4 coefficient would be slightly smaller,
$f_2 = 0.016 \pm 0.039\ {\rm (stat.)}$, but compatible with the above
result (\ref{eq:f2}) within uncertainties (with similar systematic
errors as in Eqs.~(\ref{eq:a0}) and (\ref{eq:f2})).

In earlier phenomenological analyses \cite{JM,KAO} larger values of
$f_2$ were obtained.
Using the SLAC-E143 data \cite{SLAC-E143}, Ref.~\cite{JM} found
$f_2 = 0.10 \pm 0.05$, while Ref.~\cite{KAO} used in addition
HERMES \cite{HERMES} and CLAS \cite{CLAS} data to extract a value
$f_2 = 0.15-0.18$.
In both cases, however, the $1/Q^4$ corrections in Eq.~(\ref{eq:M1twist})
were not included, which we find important even for $Q^2 \sim 1$~GeV$^2$.
This is particularly relevant for the analysis in Ref.~\cite{KAO}, which
assumes that the higher twists are dominated by the $1/Q^2$ term already
at $Q^2 \sim 0.5$~GeV$^2$.
If one were to redo the present analysis with a 1-parameter fit as
in Refs.~\cite{JM,KAO}, the statistical error on $f_2^p$ from the
$Q^2 > 1$~GeV$^2$ data would be $\sim 0.005$, which is 4--5 times
smaller than that in the 2-parameter fit, and gives a 2--3 smaller
combined statistical and systematic uncertainty than in Ref.~\cite{JM}.
The result of the present work (\ref{eq:f2}) therefore represents a
significant improvement over the earlier analyses.

From the extracted $f_2$ values one can calculate the contribution
of the collective color electric and magnetic fields to the spin of
the proton.
These are given by \cite{MANK,JI_CHI}
\begin{eqnarray}
\chi_E
&=& {2 \over 3} \left( 2 d_2\ +\ f_2 \right)\ ,		\\
\chi_B
&=& {1 \over 3} \left( 4 d_2\ -\ f_2 \right)\ ,
\end{eqnarray}
where $d_2$ is given by the matrix element of the twist-3 operator
$\bar\psi ( \widetilde G^{\mu\nu} \gamma^\alpha
	  + \widetilde G^{\mu\alpha} \gamma^\nu ) \psi$,
and can be extracted from the second moments of $g_1$ and $g_2$,
\begin{eqnarray} 
\label{eq:d2}
\hspace*{-0.5cm}
d_2(Q^2)
&=& \int_0^1 dx\ x^2 \left[ 2 g_1(x,Q^2)\ +\ 3 g_2(x,Q^2) \right] ,
\end{eqnarray}
as determined from the present analysis.
We find, however, that its leading twist component is negligible,
and consistent with zero.

Combining the extracted $f_2$ and $d_2$ values obtained from the
global analysis, we find
\begin{eqnarray}
\chi_E &=&  0.026\ \pm\ 0.015\ ({\rm stat.})\
		   \pm\ {}^{0.021}_{0.024}\ ({\rm sys.})
\label{eq:chiE}						\\
\chi_B &=& -0.013\ \mp\ 0.007\ ({\rm stat.})\ 
		   \mp\ {}^{0.010}_{0.012}\ ({\rm sys.})
\label{eq:chiB}
\end{eqnarray}
where the low-$x$ extrapolation and $\alpha_s$ uncertainties have been
folded into the total systematic error.
Since the color polarizabilities are dominated by $f_2$, the sign of
the color electric polarizability is positive, while that of the color
magnetic polarizability is negative.
The upper limit on $f_2$ in Eq.~(\ref{eq:f2}) thus yields non-zero
values for $\chi_E$ and $\chi_B$, while the lower limit gives values
which are close to zero.

These results can be compared to model calculations.
QCD sum rules generally predict negative values for the electric
polarizabilities, and slightly positive values for the magnetic
ones \cite{MANK,BBK},
$\chi_E^{\rm sum\ rule} \approx -(0.03-0.04)$ and
$\chi_B^{\rm sum\ rule} \approx 0.01-0.02$,
in contrast to the results in Eqs.~(\ref{eq:chiE}) and (\ref{eq:chiB}).
Similar results are found in the calculations based on the
instanton vacuum model \cite{INSTANTON},
$\chi_E^{\rm instanton} \approx -0.03$ and
$\chi_B^{\rm instanton} \approx 0.015$.
The MIT bag model on the other hand gives \cite{JU_G1}
$\chi_E^{\rm bag} \approx 0.03-0.05$ and
$\chi_B^{\rm bag} \approx 0.00-0.02$,
which is consistent with our findings.

More precise measurements of the structure functions at
$Q^2 \approx 1-10$~GeV$^2$, with smaller statistical and systematic
errors, would reduce the uncertainty in the extracted higher twist
coefficients, as would better knowledge of $\alpha_s$ at the relatively
low $Q^2$ values discussed here.
The present findings suggest that higher twists in the lowest moment
of the proton $g_1$ structure function are small and consistent with
zero for $Q^2 \agt 2-3$~GeV$^2$ (see also Refs.~\cite{HERMES_A1P,BFL}),
which demonstrates, perhaps surprisingly, the usefulness of the OPE
formalism at these rather low $Q^2$ values.
Higher twists are expected to play a greater role in higher moments,
which emphasize more the large-$x$ region and receive larger resonance
contributions at the same $Q^2$ \cite{BIGPAPER}.
Better determination of the $g_2$ structure function at moderate
and high $Q^2$ is also vital for the determination of the $d_2$
matrix element, as well as of the $g_1$ structure function itself.

%%%%%%%%%%%%%%%%%%%%%%%%%%%%%%%%%%%%%%%%%%%%%%%%%%%%%%%%%%%%%%%%%%%%%%%%%
We would like to thank G.~Ricco for helpful contributions.
This work was supported in part by the U.S. Department of Energy
contract \mbox{DE-AC05-84ER40150}, under which the Southeastern
Universities Research Association operates the Thomas Jefferson
National Accelerator Facility, by DOE grants \mbox{DE-FG02-96ER40960}
(Old Dominion U.) and \mbox{DE-FG02-96ER41003} (College of William and
Mary), and by NSF grant 0099540 (Hampton U.).

%%%%%%%%%%%%%%%%%%%%%%%%%%%%%%%%%%%%%%%%%%%%%%%%%%%%%%%%%%%%%%%%%%%%%%%%%

\end{document}